# Predicting Coding Effort in Projects Containing XML Code

S. Karus
University of Zurich, Switzerland
University of Tartu, Estonia
siim.karus@ut.ee

M. Dumas
University of Tartu
Estonia
marlon.dumas@ut.ee

## ABSTRACT

This paper studies the problem of predicting the coding effort for a subsequent year of development by analysing metrics extracted from project repositories, with an emphasis on projects containing XML code. The study considers thirteen open source projects and applies machine learning algorithms to generate models to predict one-year coding effort, measured in terms of lines of code added, modified and deleted. Both organisational and code metrics associated to revisions are taken into account. The results show that coding effort is highly determined by the expertise of developers while source code metrics have little effect on improving the accuracy of estimations of coding effort. The study also shows that models trained on one project are unreliable at estimating effort in other projects.

## 1. INTRODUCTION

Estimating the cost of software projects is a long-standing research problem in the software engineering field [1]. It is both highly challenging from a research perspective and highly relevant from a practical perspective. Indeed, misestimates of software project costs can result in high losses due to overdue and over-budget projects [2], [3].

It has been shown that expert judgement is highly inconsistent and thus not a reliable means of software cost estimation [4]. Accordingly, a number of algorithmic or statistical models for software cost estimation have emerged. In general, these models focus on estimating development and maintenance effort since human resources are the source of a significant portion of software development costs.

We can broadly classify development effort estimation models into *a priori models* and *evolutionary models*. The former category includes models like COCOMO [3], which aim at predicting effort based on project features that are usually known during the analysis and design phases, such as expected number of function points, size, experience and cohesion of the team, software processes and tools employed, etc. A key characteristic of these models is that they do not take into account the code base and history of the project at the time the prediction is made. In contrast, evolutionary models aim at predicting future development and maintenance effort of a project, by using metrics extracted from the project's code base, version control system, bug tracking system and other collaboration systems [5], [6], [7], [8], [9]. In other words, these methods take into account the intermediary snapshots and evolution of the software project. The study reported in this paper falls in this latter category, and focuses specifically on the use of metrics extracted from version control systems.

Another way of classifying effort estimation models is based on the measure that they aim at predicting. Models in the style of COCOMO aim at predicting man-hours. Of course, this is highly dependent on the past experience of developers in previous projects and many other organisational parameters that are exogenous to the project in question. Due to these exogenous factors, the predictive power of these models can vary across projects and careful calibration is required [3]. In contrast, other models tend to predict a measure that can be directly derived from the software product itself: namely *code churn* [10]: the sum of lines of code (LOC) added, removed or modified during a certain timeframe. While code churn does not directly measure total development or maintenance costs, it is nonetheless an indicator of coding effort during a given timeframe of a project. Additionally, code churn is an indirect indicator of the cost of design, testing, deployment and project management. The more concise are the modifications and additions, the less code is there to test, and deploy. Also, a project with a reduced number of small modifications is indicative of less major re-design decisions, compared to a project with large and frequent rewrites. Thus, estimating code churn can give an insight into total development effort – be it during the initial development phase or during the maintenance phase of a project. Furthermore, code churn has been shown to be a predictor of defects in software projects [11]. The present study deals with the prediction of three code churn metrics separately: number of added LOC, modified LOC and deleted LOC over a one-year period, which corresponds to the length of a typical long-term software development planning period and is in line with the choice of timeframe made in related studied on long-term code churn [12], [13].

Most *evolutionary* coding effort estimation models assume that coding effort can be estimated based on software design metrics (e.g. coupling and cohesion metrics [8]), or code metrics (e.g. code complexity metrics [5]). Relatively little attention has been given to estimating coding effort based on organisational and project metrics extracted from version control systems, such as developers' activity. In this respect, a recent study has shown high correlation between organisational metrics and code churn metrics in the development of the Windows operating system [6].

This and similar studies suggest that there is a wealth of information available in version control systems – beyond the code itself – that can be used to predict code churn.

In evolutionary estimation models, information extracted from the current snapshot and evolution of the project is used to predict future coding effort. Many techniques can be used for this purpose. In this paper, we rely on standard machine learning techniques. These techniques require training prior to their use. The training is done on a subset of snapshots, and the trained models are used to predict one-year code churn on other snapshots. A question that arises in this setting is whether models trained on snapshots from a project can be used to make predictions on other projects.

In light of the above, this paper addresses the following general research questions:

- RQ1. Can coding effort for the subsequent year, measured in terms of added LOC, deleted LOC, and modified LOC, be estimated based on project metrics extracted from version control systems?
- RQ2. Can a model trained on one project be used to make accurate predictions on other projects (i.e. are the main factors determining the future coding effort the same across multiple projects)?
- RQ3. Are the relationships between coding effort and project metrics fundamentally different from one project to another, or can a single unified model be used to make estimations of added, modified and removed LOC across multiple projects?

The bulk of previous research addressing the question of coding effort estimation in general and evolutionary effort estimation in particular, deal with code written in procedural and object-oriented programming languages. In the meantime, the rapid uptake of the Web has resulted in more and more software systems containing code written using the eXtensible Markup Language (XML) code and associated languages such as the Extensible Stylesheet Language Transformations (XSLT). XML-based languages are used in software projects to encode build and deployment information (e.g. Ant), configuration information, application data and data schemas (XML Schema), document transformations (XSLT), images (SVG) and other software artefacts pertaining to the presentation layer of Web applications. Statistics extracted from Ohloh.net[1] show that about 20% of open-source projects make use of XML, while less than 10% make use of Java (and the same applies to C). Yet, we are not aware of any previous study that attempts to predict coding effort from data associated with XML files in project repositories.

In this setting, this paper addresses the above research questions in the context of software projects that make use of XML. The underpinning hypothesis is that XML files in such software projects contain valuable information that can be used to build accurate code churn prediction models. In particular, the paper considers the use of machine learning algorithms to build code churn prediction models based on code and organisational metrics extracted from XML files, combined with project features extracted from the project's version control system

This paper starts with an overview of related work (Section 2), followed by the description of the method (Section 3), dataset (Section 4), metrics (Section 5), and algorithms (Section 6) employed for creating churn estimation models. The experimental results are discussed in Section 7. Finally, threats to validity are reviewed in Section 8 while conclusions and possible future directions are outlined in Section 9.

[1] Language usage values were taken on 15th August 2010 from the open-source projects database http://ohloh.net, which tracks more than 437,000 open-source projects Worldwide.

## 2. RELATED WORK

Software development effort is commonly measured in man-hours, project cost, or code churn. The first two can only be used when studying commercial software projects. In contrast, open-source public (especially community-driven) projects do not capture data about development time or project cost. The only data available for analysis of open-source projects has to come from version control systems, mailing lists, and bug tracking systems. These sources give information about developer past activity, which can be incorporated into development effort estimation models to balance the lack of other information in case of open-source software [7]. The work reported here is based on open-source project and accordingly, it aims at predicting coding effort in terms of product characteristics (churn metrics) rather than man-hours or monetary costs.

The data used for estimation also differs between studies [14]. Commonly used input data for estimation includes: analysis documents, source code/design metrics, expert judgement, and organisational/project metrics. Most algorithmic approaches base their estimations on either project metrics or code metrics but not on both. This paper studies the question of whether combining these two approaches would result in better estimations.

Zhou, et al. [8] used linear regression to investigate relations between object-oriented design metrics and maintainability in open source Java projects. This study is representative of a body of studies aimed at testing hypotheses such as "low coupling between classes in an object-oriented software leads to better maintainability measured in terms of amount of changes during the maintenance phase" [12]. Similar studies consider the use of procedural code complexity metrics to predict maintainability [5]. These previous studies focus on code metrics, while our study combines organisational and code metrics, and focuses on projects containing XML code.

Nagappan, et al. [6] used organisational metrics to estimate code churn in the development of the Windows operating system. This study addresses research questions similar to RQ1. However, the organisational metrics considered by Nagappan, et al. are only available in commercial software projects. Our study focuses on metrics that can be extracted from a project's source code management system, and therefore the study can be applied both in commercial and open-source projects. That is, we extend the idea of using organisational metrics for evolutionary effort estimation to a broader range of software projects.

In a similar vein, Pendharkar, et al [9] studied the relation between team size and software development cost (including initial development and maintenance). They uncovered a significant correlation between "active" team size and coding effort measured in terms of added, modified and deleted lines of code. Our study follows a similar line but considers a broader set of metrics.

## 3. METHOD

The aim of the study is to determine if it is possible to build statistical models to predict the future code churn of a project (specifically added LOC, deleted LOC, and modified LOC) based on metrics extracted from version control systems. Given this aim, we have a choice between hypothesising that certain relations exist between a set of input metrics and the above three code churn metrics, or to uncover such relations using exploratory analysis. In the first approach we would start with a set of hypothesis, and we would use statistical conformance testing to validate these hypotheses on the chosen dataset. However, as mentioned above, we are not aware of previous studies on possible relations between code/project metrics and code churn metrics in the context of projects containing XML code. Hence, there is little basis for formulating a priori hypotheses about such relations. Accordingly, we adopt a bottom-up approach based on data mining and exploratory data analysis.

**Table 1. Details on projects used in the study**

| Project | Common files | Revs | Max Files | Devs | Years |
|---|---|---|---|---|---|
| **Commons**<br>http://www.wso2.org/ | no ext.&java (31%),<br>**XML (11%)** | 1711 | 1517 | 37 | 3 |
| **Dia**<br>http://www.gnome.org/projects/dia/ | shape (28%),<br>xpm (14%),<br>c&png (13%) | 3571 | 632 | 139 | 11 |
| **Docbook**<br>http://docbook.sourceforge.net/ | **XML (34%)**,<br>gen (25%),<br>no ext. (14%) | 7136 | 3028 | 28 | 8 |
| **Docbook2X**<br>http://docbook2x.sourceforge.net/ | **XSL (36%)**,<br>**XML (17%)**,<br>no ext. (11%) | 1081 | 159 | 2 | 8 |
| **Esb**<br>http://www.wso2.org/ | **XML (27%)**,<br>java (18%),<br>**XSL (14%)** | 1070 | 546 | 10 | 3 |
| **eXist**<br>http://exist.sourceforge.net/ | java (60%),<br>**XML (7%)**,<br>jar (6%) | 6300 | 3329 | 31 | 7 |
| **feedparser-read-only**<br>http://www.feedparser.org/ | **XML (78%)**,<br>HTML (19%),<br>no ext. (1%) | 227 | 1287 | 4 | 3 |
| **Gnome-doc-utils**<br>http://live.gnome.org/GnomeDocUtils | **XML (23%)**,<br>po (19%),<br>**XSL**& no ext. **(16%)** | 968 | 121 | 115 | 5 |
| **Groovy**<br>http://groovy.codehaus.org/ | java (40%),<br>groovy (29%),<br>jar (7%) | 6920 | 2549 | 53 | 7 |
| **Tei**<br>http://tei.sourceforge.net/ | odd (41%),<br>**XML (34%)**,<br>**XSL (12%)** | 4318 | 1633 | 12 | 6 |
| **Valgrind**<br>http://valgrind.org/ | c (25%),<br>exp (19%),<br>vgtest (14%) | 4090 | 1000 | 19 | 2 |
| **Wsas**<br>http://www.wso2.org/ | java (31%),<br>**XML (18%)**,<br>no ext. (17%) | 1180 | 823 | 22 | 3 |
| **Wsf**<br>http://www.wso2.org/ | c (15%),<br>h (14%),<br>no ext. (13%) | 1405 | 2285 | 21 | 3 |

The adopted data mining approach comprises the following steps:

1. Data pre-processing: choice of prediction targets and proposition of input features (attributes that might influence the value we need to predict), data gathering, normalisation, and cleansing.
2. Learning: choice of data mining algorithms and application of these algorithms.
3. Results validation: evaluation of models fit using standard statistical techniques.

This data mining approach allows us to identify interference between input features and the prediction target and in doing so they uncover the existence of a predictive model. However, the data mining approach itself does not allow us to explain the cause of the interference. In order to compensate for this shortcoming, exploratory data analysis was used in order to gain an understanding of the models created by the data mining algorithms.

Compared to a conformance testing approach, data mining and exploratory data analysis offer the benefit of not requiring an a priori specific model to test. The aim of exploratory data analysis is to propose models that can then be conformance-tested. Moreover, data mining and exploratory data analysis can uncover non-intuitive relations. In fact, the results of this study show, among other things, that there are no generally-applicable straightforward models to estimate coding effort – that is, linear models based on one or a very small number of interactions between input features. The models with better predictive power uncovered in the study involve a non-trivial number of interactions between input features.

# 4. DATASET

Eight Open Source Software Project repositories with XSLT and XML code were used to train the models:

- WSO2 commons
- WSO2 Wsas
- WSO2 Esb
- Docbook
- Docbook2X
- Exist
- Gnome-doc-utils
- Tei

Additionally, five projects were used only for testing:

- Feedparser-read-only
- WSO2 wsf
- Dia
- Groovy
- Valgrind

These projects were chosen so that they would represent different types of software systems. For example, WSO2 is an enterprise service bus type of a project, docbook, docbook2X and gnome-doc-utils are used for documentation formatting, dia for graph drawing and exist, tei, groovy and valgrind are software development tools. More than 118,000 file revisions (of more than 24,000 files) were used for data preparation, analysis, model creation, and testing. It is noteworthy that project "gnome-doc-utils" did not have any lines of code modified or removed and did not therefore produce models for predicting modified or removed LOC. An overview of the project repositories used is given in Table 1.

A common term for long-term code churn prediction is a year ([12], [13]), which is used in this study as well. Added, modified and removed lines of code for each file revision was calculated based on GNU diff output. Yearly modified, added and removed LOC are considered to be the sums of corresponding file revision operation measurements during the year following the date of commit of the revision ("cumulative yearly added LOC", "cumulative yearly deleted LOC", and "cumulative yearly modified LOC").

As XML is also used for storing project information (e.g. ant build scripts), we did have a look into the types of XML files used in the projects. It turned out that out of all XML files 13% were project definitions and ant build files while the rest were mostly project specific files (that is, they used namespaces defined in the project itself). All projects had XML files that were neither build nor project definition files.

The projects studied had life spans from 2 to 11 years and some of these are still active. The data was collected during spring-summer of 2009. For each project, all revisions over the entire project's lifetime were extracted. We could have partitioned the dataset into one-year periods (e.g. calendar years). However, this would have reduced the amount of one-year periods available for training and validation of the algorithms. Instead, we decided to compute for each project revision, the code churn over the one-year period starting from the date of the commit of the revision. In this way, there were as many predictions as the number of commits. Of course, it was not possible to compute one-year-forward code churn metrics for revisions with commit dates within one year of the date of extraction of the dataset, so these revisions were not included in the set of revisions for training and validation of the learning algorithms (though they were used to calculate one-year-forward code churn of previous revisions).

# 5. FEATURE SELECTION

In order to train models for estimation, features to base the estimations on, need to be selected. In our study, we selected two sets of features independently: organisational/project metrics extracted from version control systems, and code metrics extracted from XML and XSLT files in a given snapshot of the project.

## 5.1 Project features

Project features describe the project team and the familiarity of the team with some of the most popular technologies/languages. In addition, project size (in number of files) and project age/maturity (in revisions) were taken into account. Number of file extensions was used as an indicator of technologies used in the project both historically and currently. The committers' previous activities were also included in the feature set as it can be assumed that the developer making the commit is still active whilst others might no longer be active participants.

For each file revision in the version control system, the following metrics were calculated:

- Number of developers, who have made commits to version control system before and including the time of the commit of the revision at hand (DevelopersOnProjectToDate)
- Number of developers, who have made commits to XSL files in version control system before and including the time of the commit of the revision at hand (XSLDevelopersOnProjectToDate)
- Number of developers, who have made commits to XML files in version control system before and including the time of the commit of the revision at hand (XMLDevelopersOnProject)
- Number of developers, who have made commits to Java files in version control system before and including the time of the commit of the revision at hand (JavaDevelopersOnProject)
- Number of developers, who have made commits to HTML files in version control system before and including the time of the commit of the revision at hand (HTMLDevelopersOnProject)
- Number of developers, who have made commits to C files in version control system before and including the time of the commit of the revision at hand (CDevelopersOnProject)

- Number of developers, who have made commits to png, jpg and gif files in version control system before and including the time of the commit of the revision at hand (GraphicsDevelopersOnProject)
- Number of previous commits made by the committing developer (NumberOfDeveloperPreviousCommits)
- Number of previous commits made by the committing developer to XSL files (NumberOfDeveloperPreviousXSLCommits)
- Number of previous commits made to project (NumberOfRevisions)
- Current number of files in project (NumberOf Files)
- Current number of different file extensions used in the project (NumberOfFileExtensions)
- Number of all file extensions used in the project to date (NumberOfHistoricFileExtensions)
- Cumulative yearly added LOC (prediction target, used only for validation and training)
- Cumulative yearly modified LOC (prediction target, used only for validation and training)
- Cumulative yearly deleted LOC (prediction target, used only for validation and training)

The number of developers by area of expertise was calculated by counting committers who made commits to files with the corresponding file extension. For example, a Java developer is defined as one who has made at least one commit to a .java file, while a C developer is one who has made at least one commit to a .c or .h file, a Graphics developer is one with commit to .gif, .png or .jpg file, and an XML/XSL/HTML developer is one with commit to .xml, .xsl (or .xslt) or .html (or .htm) file, respectively. We only considered five areas of expertise (Java, C, Graphics, XML, XSL and HTML) because these were the only file types that appeared in abundant amounts in the studied data set.

## 5.2 Code features

Maintainability estimation studies on Object-Oriented and structural languages commonly use lines of code, number of classes, inheritance metrics and different program flow based complexity metrics. XML and XSLT, however, have neither classes nor inheritance. Additionally, program flow graphs of XSLT transformations are difficult to construct due to the non-procedural nature of XSLT [15]. Thus, new metrics need to be considered for XML and XSLT.

On the other hand, because of its structure, it is straightforward to define "count metrics" on XML (e.g. number of elements and number of attributes). Furthermore, elements have different types (i.e. names) and thus we can potentially define one count metric per element. However, the set of possible XML element or attribute names is infinite, making the definition of one count metric per possible type of XML element impractical. Thus, we chose to count only occurrences of each type of XSLT 1.0 element, as well as the total number of elements and attributes in an XML file not belonging to the XSLT namespace. We did not count occurrences of elements that can be present only once per transformation (e.g. <transformation>, <stylesheet>, <output>).

XSLT uses XPath expressions for matching, selecting and testing input data. These expressions can only be present in the XSLT attributes "match", "select", "test" and in attributes outside the XSLT namespace enclosed between curly braces (these are called "inline expressions"). Inline XPath expressions can only be used for selecting data. As these expressions denote decision points in XSLT, counting them gives us information on the flow of transformations.

It is important to differentiate between two types of XPath expressions: simple expressions identifying specific elements or attributes by their name and namespace versus complex expressions identifying wider ranges of nodes in input documents. Complex expressions can be written using wildcards or using function calls. Simple expressions are those that do contain neither a wildcard nor a function call. By counting separately simple and complex expressions we can get an indication of the complexity of the transformation.

In total, 61 primary metrics (also called *features*) were collected for each project snapshot, that is, for each commit found in each project repository. The 61 metrics include:

- count of XML nodes,
- count of different types of XML nodes (4 metrics),
- count of each type of XSLT element (28 metrics),
- count of XSL output literals,
- count of elements in XSL target namespace,
- count of direct children of the root element,
- count of XSL global parameters and variables (2 metrics),
- count of inline expressions,
- count of XSL attributes that contain test expressions ("select", "match", and "test") for each expression type ("simple", "complex due to wildcard usage", and "complex due to function call usage") and total number of select, test and match elements (12 metrics),
- average number of child elements of XSLT "message" elements,
- sum of XML attribute and element type nodes,
- count of attributes and elements inside XSLT "variable" element,
- count of attributes and elements inside XSLT "param" element,
- count of attributes and elements inside XSLT "message" element,
- count of XSLT output attributes and elements (sum of XSLT "element" elements, XSLT "attribute" elements, and XML elements and attributes in target namespace,
- count of "complex" expressions by attribute ("select", "match", "test") and total number of complex expressions in the file (4 metrics).

# 6. TRAINING ALGORITHMS

Two different algorithms were used for training models for coding effort estimation:

- Neural Networks (NN) – Back-Propagated Delta rule network with three layers
- Decision Trees (DT) with regression equations on the leaves

The neural networks algorithm is capable of identifying various complex relations between input features and prediction target. Due to this ability, Neural Networks are commonly used to build models for estimation of features that have complex relations with other features. As such, Neural Networks can be used to build highly accurate estimation models in cases where relations between features are not known well or involve many interactions between features. On the other hand, the performance of Neural Networks algorithm and the models built using the algorithm is highly dependent on the amount of training data available.

**Table 2. Mean and median correlations and normalised MAE and RMSD for models trained only on organisational metrics.**

| | Operation | Pearson | | Kendall | | NMAE | | NRMSD | |
|---|---|---|---|---|---|---|---|---|---|
| | | Mean | Median | Mean | Median | Mean | Median | Mean | Median |
| DT | Added | 0.9101 | 0.9216 | 0.7308 | 0.7064 | 0.1509 | 0.1078 | 0.1050 | 0.0964 |
| | Removed | 0.9301 | 0.9443 | 0.7759 | 0.7978 | 0.1556 | 0.1073 | 0.1051 | 0.1026 |
| | Modified | 0.9287 | 0.9420 | 0.7877 | 0.7785 | 0.1471 | 0.0939 | 0.1008 | 0.1007 |
| NN | Added | 0.8169 | 0.9147 | 0.6695 | 0.7140 | 0.0336 | 0.0146 | 0.1526 | 0.1192 |
| | Removed | 0.9106 | 0.9502 | 0.7253 | 0.8048 | 0.0329 | 0.0172 | 0.1336 | 0.1233 |
| | Modified | 0.9147 | 0.9502 | 0.7297 | 0.7424 | 0.0354 | 0.0199 | 0.1332 | 0.1237 |

Decision Trees allow one to build simpler and more explainable models. This also makes models built by Decision Trees algorithm easier to infer and interpret. The performance of the algorithm and models built by it are also less dependent on the amount of training data used. However, Decision Trees algorithm cannot handle complex relations well. Thus, Decision Trees (with regression equations) and Neural Networks are somewhat complementary. This is confirmed by a study by MacDonell & Gray [16] who empirically tested various learning techniques and found linear regression (with removal of outliers) and Neural Networks the best learning methods for estimations on software projects.

Microsoft SQL Server 2008 Analysis Services was used to train and test the models. The dataset was split into training (70%) and test set (30%) using random sampling and the rules outlined in section 4. To aid the Neural Networks algorithm, added, removed and modified LOC values were normalised for training and testing of Neural Networks trained models. Cross-validation (3- and 4-fold) on subsets of up to 7000 entities was used to verify that the algorithms behaved consistently with low chance of over-fitting (identification of random patterns). Cross-validation on full dataset or with more folds (e.g. 10-fold) failed due to database limitations.

# 7. RESULTS

The training resulted in fairly complex models. However, tests made on testing data set do give information about their usefulness. The tests used to validate the models were:

- Pearson Correlation coefficient – indicates linear correlation between actual and predicted values (higher value is better)
- Kendall Correlation coefficient – indicates rank based correlation between actual and predicted values (higher value is better)
- Mean Absolute Error (MAE) – Mean difference between actual and predicted values (lower value is better)
- Normalised Mean Absolute Error (NMAE) – Normalised MAE (MAE divided by the mean of actual values)
- Root Mean Square Deviation (RMSD) – measures differences between predicted values and actual values. Compared to MAE, RMSD is more sensitive to high errors (lower value is better)
- Normalised Root Mean Square Deviation (NRMSD) – min-max normalised RMSD

In the following discussion only normalised values are mentioned as absolute values are not comparable between projects and algorithms used.

## 7.1 RQ1: Project-Specific Models

The first research question is whether we can construct prediction models based on code metrics or organisational metrics extracted from version control systems. To this end, we constructed models for each project using organisational metrics alone, code metrics alone and a combination of both.

The means and medians of validation measures for models trained on organisational metrics alone are given in Table 2. This table shows that the generated models have a reasonable ability to estimate coding effort. Most models had high predictive power, however, the mean values are penalised by the low predictive power of some models (as shown by significantly better median values in most cases). Namely models trained on “docbook2X” project had rather low predictive power with Pearson correlations around 0.82, Kendall correlations around 0.63, NMAE around 0.28 and NRMSD around 0.16 among models trained using the Decision Trees algorithm. (The corresponding Neural Networks models were slightly better with Pearson around 0.83, Kendall around 0.55, NMAE around 0.08 and NMRSD around 0.10).

For models trained on both organisational and code metrics, the validation measures show only a slight difference in predictive power on average – improving the performance in some cases, but deteriorating it in other cases. For models trained using Decision Trees on all metrics, Pearson correlations dropped by 0.0343 on average and Kendall correlations by 0.0568 on average while NMAE values increased by 0.0547 on average and NRMSD values increased by 0.0327 on average compared to models trained only on organisational metrics. Estimations of models trained on the Exist project clearly improved when code metrics were excluded from the training of the Decision Trees models. On the other hand, models trained using Neural Networks were consistently worse when code metrics were excluded, and Kendall correlation slightly increased on average with the addition of code metrics. This can be explained by the fact that, in the majority of the cases, Neural Networks models had lower performance when trained on only organisational metrics compared to corresponding models trained using Decision Trees algorithm. It is also noteworthy that with low amounts of training data, the inclusion of code metrics improved the accuracy of Neural Networks models more than it did when high amounts (more than 2500 training cases) of training data was employed.

Some models trained on all metrics (organizational plus code metrics) displayed an interesting characteristic of identifying organisational metrics either as having low-influence dependencies or sometimes ignoring them completely. Still, the differences in terms of validation metrics compared to models built on only organizational metrics were small. For example, models trained using Decision Trees on project "esb" had Pearson correlation higher than 0.99, Kendall correlation above 0.94, NMAE less than 0.02 and NMRSD less than 0.05 on project "esb" when all features were used during training, compared to Pearson correlations higher than 0.96, Kendall correlations higher than 0.89, NMAE less than 0.06 and NMRSD less than 0.08 for models trained only on organisational metrics (reducing error at 90% confidence by 2-7%). Similarly models trained on project "tei" improved from Pearson correlations from 0.92 to 0.96, Kendall correlations from 0.75 to 0.84, NMAE from 0.13 to 0.08, NRMSD from 0.12 to 0.09 (reducing error at 90% confidence by 2-6%). This could be caused by correlations between organisational and code metrics.

**Table 3. Main dependencies of models trained using Decision Trees on projects Esb and Tei on organizational metrics.**

| esb | Tei |
|---|---|
| Developers on Project To Date | Number of Files (only for LOC Modified and Removed, split at 1375) |
| Number of Revisions (only for LOC Added) | Developers on Project To Date (only for LOC Added) |
| Number of Files (only for LOC Modified and Removed) | XSL Developers On Project To Date |
| XSL Developers On Project To Date (4th for LOC Modified and Removed) | Developer Previous XSL Commits |
| Developer Previous XSL Commits | Developer Previous Commits |
| Developer Previous Commits | Developers on Project To Date (only for LOC Removed) |

Overall, the results show that organisational features are important drivers of coding effort. Thus, it is possible to train highly accurate models for coding effort estimation based only on organisational features. The inclusion of code metrics improves the predictive power of the models in most cases, but not in a consistent manner.

## 7.2 RQ2: Applying Models across Projects

In order to address RQ2, we tested whether models trained on only a given project can give good estimations on other projects.

The test results show that models trained on single project are not generalisable. Estimations on all the projects did not have any predictive power – the correlations with actual values were close to zero or even negative. Consistently positive correlations were present only for the models trained on project Exist using Decision Trees algorithm showed consistently low positive correlations (Pearson ranged from 0.1073 to 0.1441 and Kendall from 0.2241 to 0.2681) and models trained on project Docbook with Neural Networks algorithms (Pearson ranged from 0.1276 to 0.3364 and Kendall ranged from 0.1990 to 0.3365).

When looking at predictions made by models trained using Decision Trees algorithm per project, it was quite clear that projects "docbook" and "tei" were outliers. In fact, estimations made by most models had significant negative or non-existing correlations with actual values (Pearson mostly between -0.35 and -0.84, Kendall mostly between -0.25 and -0.57). Models trained on projects "commons", "esb", "exist", "tei" and "wsas" were good at estimating removed or modified LOC for project "commons" (Kendall correlation from 0.4115 to 0.8763), however, model trained on project "exist" was the only one but the model trained on project "commons" itself to have high positive correlations (Pearson 0.6200, Kendall 0.5441) with "added LOC" for project "commons". On overall, project "esb" was the easiest to make estimations on – only models trained on projects "commons", "exist" and "tei" made estimations with low correlations (Pearson less than 0.55, Kendall less than 0.37) toward actual values. It is interesting to note that model trained on project "tei" made estimations with strong negative correlations (Pearson less than -0.70, Kendall less than -0.15 (-0.69 in case of "Added LOC")) with actual values for project "esb". This strong distinction implies that these projects have distinctly different drivers for coding effort. Despite these differences, the main dependencies of coding effort were similar for both of the models as can be seen in Table 3.

Figure 1 shows the actual decision trees produced by on "esb" and "tei" projects. It can be seen that despite the dependencies for modified and removed LOC are similar, the way these dependencies affect modified and removed LOC is very different. For one, the number of developers on project has negative effect on modified and removed LOC in project "esb", but positive effect in project "tei". In fact, while studying the models created by Decision Trees algorithm, we were not able to find any feature that had the same kind of effect (either always positive or always negative) in all projects. Thus, the features studied are not suitable for generalised models on their own, implying the need to include some other additional features or non-linear and non-sigmoid transformations of features or analysis of interactions between features in order to build models that could explain drivers of coding effort for all projects in general.

The study of NMAE values of models trained using Decision Trees algorithm, however, shows that despite high correlations, the errors of estimations on other projects are enormous (ranging from 0.31 to 391.27), making the models useless for predictions on other projects. Only predictions on project "esb" by models trained on project "docbook" are of some use with NMAE from 0.22 to 0.45. NMRSD values confirm that in general, a model trained on one project fails to give good estimates on other projects (NMRSD ranging from 0.22 to 141.59). Whilst models trained on all features performed slightly better, the errors in models trained using Decision Trees algorithm were still too high to make the models useful on projects other than the one used for training.

The fact that having more developers could reduce the coding effort (number of modified, added or removed lines of code) intrigued us as it is highly counter-intuitive. We made an additional test to see whether high developer turnaround could have caused this phenomenon by counting the number of active developers (developers who have made commits in the past and will make future commits during the period studied) in addition to all developers. The test showed that even though the number of active developers was considered an important influencer for some models, the number of total developers had always similar influence (the multipliers were differed less than 5%). Thus, high developer turnaround was not confirmed as the cause of the phenomenon.

Another possible cause of the phenomenon is more diverse code ownership. That is, developers are reluctant to modify other developer's source code or the diversity of developers makes the source code more difficult to comprehend. This hypothesis was not tested as part of this study and would still need to be put on test.

Finally, we trained models using Neural Networks on a given project and tested them on all other projects. Models trained using neural networks predicted code churn on other projects better than models trained using Decision Trees, occasionally having Pearson correlation higher than 0.9, Kendall correlation higher than 0.8, NMAE lower than 0.05 and NMRSD lower than 0.2 for estimations on other projects. These occasions were exceptions and in the vast majority of cases the models did not have significant predictive power on projects they were not trained on.

Thus, we conclude that models trained on the repository of one project are generally able when used to predict code churn on other projects.

## 7.3 RQ3: Existence of a Unified Model

To test the viability of using a single model, as opposed to training models on a project-specific basis, we trained models using data from all the projects. Models trained using Decision Trees had a reasonable Pearson correlation (higher than 0.69) and NRMSD (lower than 0.14). However, NMAE values of 0.4372, 0.6670, and 0.6227 (for estimations on LOC added, modified and removed) showed that the model was not reliable for general purpose. Models trained using Neural Networks algorithm were significantly better having Pearson correlations of 0.7131, 0.7431, and 0.7310; NMAE values 0.0802, 0.1222, and 0.1182; and NRMSD values 0.0893, 0.1264, and 0.1278. However, Kendall correlations were weak: 0.4674, 0.4077, and 0.3574. Thus, these models can be used with some success for estimations on different projects. Per project analysis confirms that models trained using Neural Networks algorithm are good at making estimations on several projects, which had the most influence during training due to their higher number of training cases. As such the models have good predictive power on projects "docbook", "docbook2X", "gnome-doc-utils", and "tei" while failing with other projects.

As projects belonging to one larger complex project are likely to have the same influencers, it is of some interest to see, whether models trained on complex project WSO2 subprojects "common", "esb" and "wsas" are usable on each other. It turns out they are not usable for predictions on each other as NMAE value for estimations on other projects is around one or even higher, meaning that average error for these predictions is close to or higher than the range of actual values.

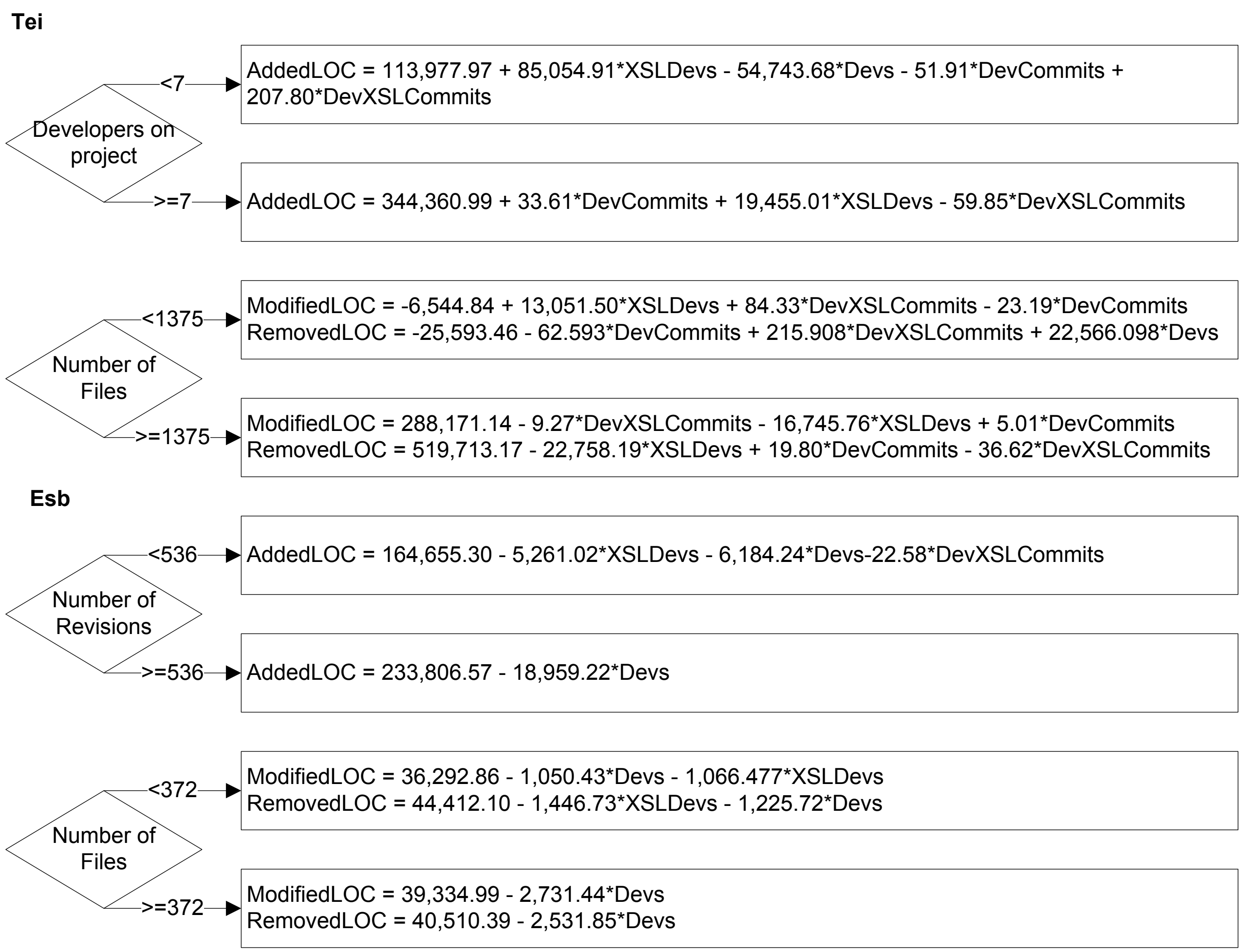

**Figure 1**. **Models trained using Decision Trees algorithm on projects "esb" and "tei" on only organisational metrics.**

We also established that similar to models trained on only organisational features, models trained on all metrics are not generalisable. To this end we studied correlations between different models' predictions and discovered that it is common for a model trained on one project to give estimations that have strong negative correlations (sometimes even with correlation coefficient values below -0.90) with actual values. This means that the code metrics used are also not sufficient for building generalisable models.

# 8. THREATS TO VALIDITY

## 8.1 External threats

One external threat to our studies is their generalisability. This threat only affects RQ1, which in its nature is internally focused. It is possible that there are projects for which coding effort estimation models cannot be successfully trained using the algorithms discussed in this study. However, all of the cases studied (some on more than 100 revisions) did yield statistically significant and useful models. As it has been suggested that 4-10 samples are usually enough to find any counter-examples [17], the likeliness of a project not having a coding effort estimation model based on organisational features is low. The diversity of the projects studied and the models achieved support the generalisability of the conclusions. There might still be some common nominators among the projects, which are the cause of the presence of these relatively simple relations.

For one, all the projects studied were open source non-commercial projects, which may differ substantially from closed source commercial projects, which might display other characteristics and be less influenced by the organisational features of the project. Another common treat among the projects studied was their rather high amount of XML and XSL code, which would definitely influence the models, especially as many of the features studied were XSL specific. Nevertheless, similar organisational metrics can be constructed for other technologies used in projects to better reflect the project and be applicable as model source metrics. The main conclusion that organisational metrics are good input features for models estimating coding effort in XML rich open source non-commercial projects is not influenced by these threats, but generalisability to larger set of projects would need further study even though the implications of these models being present in other open source non-commercial projects are strong.

The substantial differences between the two models discussed in greater detail, do raise the question of generalisability of case studies of software projects. This is especially an issue for estimation case studies where generalisability is often overlooked as not an important issue. The lack of generalisability of models based on software metrics is further encourages by the findings on reference models, which used code metrics in addition to organisational metrics, and were still not generalisable. The model trained on all the projects is an example that even if multiple projects are studied and a single model with good overall characteristics can be built on them, it is still important to validate the model per project basis to rule out erroneous generalisations.

## 8.2 Internal threats

The study of models generated is subject to the threat of not having considered all possible influencers and being missing some input features or feature transformations or interactions that could be the key to creating a single unified model that could be applicable on all projects with very good accuracy. Nevertheless, the study of actual models shows that even if there are some features independent of the features studied, linear models would still be impossible to create with the inclusion of them due to different regressions with opposite correlations with features were identified for different feature value ranges. As such, one would need to apply transformations on the features or take into account the interactions between different features in attempt to create unified models that could be successfully used for multiple projects. This would, however require the use of more complex training algorithms or additional data pre-processing.

One might also consider the option that code features and organisational features are correlating or derivable from each other. Even if this were the case, it would not discredit using models trained on all features as reference models as the assumption of the independence of code and organisational metrics is never assumed for that model.

## 8.3 Construct validity

The algorithms used in the study are not capable of identifying complex regressions – Decision Trees algorithm only identifies linear regressions and Neural Networks algorithm only combines sigmoid functions. Thus, it is possible there are relations not identified that could yield models with greater accuracy and generalisability. This does not influence the conclusions that the algorithms studied do not generate generalisable models, which has been shown by presenting counter examples (projects with contradicting models).

Pearson's correlation coefficient is generally used under the assumption of normal distribution of values, which does not necessarily hold on the datasets in the study. However, only the general form and interpretation of Pearson's correlation coefficient for testing correlation between two value sets is used in the study. We also make use of Kendall correlation, which is not affected by the distribution of values – only their order is relevant for the computation and interpretation.

Models with low error can have low Kendall correlation in case of value distributions with strong peaks. That is why MAE and RMSD are used along with Kendall and Pearson correlations to validate the models' predictive power. High Kendall correlation with high MAE and RMSD is a sign that there exists a function of the predicted value, which gives very accurate estimations of actual values (i.e. there is a missing step in the model).

# 9. CONCLUSIONS AND FUTURE WORK

We have shown that, in the context of projects with XML code, Decision Trees can be successfully used to train models to predict coding effort for the subsequent year of a project based on organisational metrics extracted from version control systems (RQ1). The inclusion of code metrics does not always improve the performance of models. The results also show that different projects have different drivers of coding effort as models trained for specific projects perform poorly when used to predict coding effort on other projects. Thus, a model trained on one specific project cannot be used for estimations on other projects (RQ2). Even models trained on data from all projects performed poorly when making effort predictions on some of the projects, showing that a single uniform model cannot be created based on the organisational used in this study (RQ3).

A more in-depth study of project-specific models aimed at identifying predictive patterns and project characteristics that affect the predictive power of these models, is a possible avenue for future work. Such a study would lead to a deeper understanding of the drivers of coding effort and possibly to the identification of early signs of code maintenance and extensibility issues. It would be of special interest to reverse the models in order to give models for optimising software project team composition and/or choice of platform and technologies used or the planned software feature-set.

Another avenue for future work is the construction of additional features to be given as input to the models. For example Smith, et al. made use of genetic algorithms to construct and select features [18]. Their approach could be used to successfully construct complex features that might lead to better models by solving and explaining non-linear relations. This could also lead to the discovery of some of the general rules and practices that affect coding effort.

It might be possible to extract additional metrics from the project snapshots. Also, additional features could be extracted from other sources, such as bug tracking systems. Those additional features might prove to have significant influence on coding effort. Identifying metrics with greater influence can result in significantly improved models.

## 10. ACKNOWLEDGEMENTS

This research was started during a visit of the first author to the software evolution and architecture lab at University of Zurich (visit funded by the Scientific Exchange Programme NMS-CH). We thank Harald Gall and the members of his group for their valuable advice. The work is also funded by ERDF via the Estonian Centre of Excellence in Computer Science.

## 11. REFERENCES

[1] B. Boehm, C. Abts, and S. Chulani, "Software development cost estimation approaches — A survey," *Annals of Software Engineering*, vol. 10, no. 1, pp. 177-205, 2000.

[2] T. C. Jones, *Estimating software costs*. Hightstown, NJ, USA: McGraw-Hill, Inc., 1998.

[3] B. W. Boehm et al., *Software Cost Estimation with Cocomo II*. Upper Saddle River, NJ, USA: Prentice Hall PTR, 2000.

[4] S. Grimstad and M. Jųrgensen, "Inconsistency of expert judgment-based estimates of software development effort," *Journal of Systems and Software*, vol. 80, no. 11, pp. 1770 - 1777, 2007.

[5] J. F. Ramil and M. M. Lehman, "Metrics of Software Evolution as Effort Predictors - A Case Study," in *ICSM '00: Proceedings of the International Conference on Software Maintenance (ICSM'00)*, Washington, DC, USA, 2000, p. 163.

[6] N. Nagappan, B. Murphy, and V. Basili, "The influence of organizational structure on software quality: an empirical case study," in *ICSE '08: Proceedings of the 30th international conference on Software engineering*, Leipzig, Germany, 2008, pp. 521-530.

[7] J. J. Amor, G. Robles, and J. M. Gonzalez-Barahona, "Effort estimation by characterizing developer activity," in *EDSER '06: Proceedings of the 2006 international workshop on Economics driven software engineering research*, Shanghai, China, 2006, pp. 3--6.

[8] Y. Zhou and B. Xu, "Predicting the maintainability of open source software using design metrics," *Wuhan University Journal of Natural Sciences*, vol. 13, no. 1, pp. 14-20, February 2008.

[9] P. C. Pendharkar and J. A. Rodger, "The relationship between software development team size and software development cost," *Communications of the ACM*, vol. 52, no. 1, pp. 141-144, January 2009, doi = http://doi.acm.org/10.1145/1435417.1435449.

[10] J. Munson and S. Elbaum, "Code churn: a measure for estimating the impact of code change," in *Proceedings. International Conference on Software Maintenance, 1998.*, 1998, pp. 24-31.

[11] N. Nagappan and T. Ball, "Use of relative code churn measures to predict system defect density," in *ICSE '05: Proceedings of the 27th international conference on Software engineering*, St. Louis, MO, USA, 2005, pp. 284-292.

[12] C. v. Koten and A. Gray, "An application of Bayesian network for predicting object-oriented software maintainability," *Information and Software Technology*, vol. 1, no. 48, pp. 59-67, January 2006.

[13] M. M. T. Thwin and T.-S. Quah, "Application of neural networks for software quality prediction using object-oriented metrics," *Journal of Systems and Software*, vol. 76, no. 2, pp. 147-156, May 2005.

[14] M. Jorgensen and M. Shepperd, "A Systematic Review of Software Development Cost Estimation Studies," *Software Engineering, IEEE Transactions on*, vol. 33, no. 1, pp. 33 -53, January 2007.

[15] M. Benedikt, W. Fan, and F. Geerts, "XPath satisfiability in the presence of DTDs," *Journal of the ACM*, vol. 55, no. 2, pp. 1-79, May 2008.

[16] S. G. MacDonell and A. R. Gray, "Alternatives to Regression Models for Estimating Software Projects," in *Proceedinings of the IFPUG Fall Conference*, Dallas, 1996, pp. 279.1-279.15.

[17] K. M. Eisenhardt, "Building Theories from Case Study Research," *The Academy of Management Review*, vol. 14, no. 4, pp. 532-550, October 1989.

[18] M. G. Smith and L. Bull, "Feature Construction and Selection Using Genetic Programming and a Genetic Algorithm ," in *Lecture Notes in Computer Science: Genetic Programming*.: Springer Berlin / Heidelberg, 2003, vol. 2610, pp. 93-100.